\documentclass{mem}
\usepackage{natbib}\usepackage{txfonts}\usepackage{balance}
\usepackage{graphicx}
\usepackage[a4paper]{hyperref}
\idline{75}{282}
\begin{document}
\def\teff{$T\rm_{eff }$}
\def\kms{$\mathrm {km s}^{-1}$}

\title{Constraining a possible time-variation of the gravitational
constant through ``gravitochemical heating'' of neutron stars}

\subtitle{}

\author{
Andreas Reisenegger\inst{1,\ast} \and Paula Jofr\'e\inst{2} \and
Rodrigo Fern\'andez\inst{3,4}
          }

  \offprints{A. Reisenegger}

\institute{Departamento de Astronom{\'\i}a y Astrof{\'\i}sica,
Pontificia Universidad Cat\'olica de Chile, Av. Vicu\~na Mackenna 4860, Macul, Santiago, Chile \and
Max-Planck-Institut f\"ur Astrophysik, Karl-Schwarzschild-Str. 1, Garching, Germany \and
Department of Astronomy and Astrophysics, University of Toronto, ON M5S 3H4, Toronto, Canada \and
Present Address: Institute for Advanced Study, Princeton, NJ 08540, USA. \\
$^\ast$ \email{areisene@astro.puc.cl} }

\authorrunning{Reisenegger et al.}

\titlerunning{Constraining $\dot G$ through heating of neutron stars}

\abstract{A hypothetical time-variation of the gravitational
constant $G$ would make neutron stars expand or contract, so the
matter in their interiors would depart from beta equilibrium. This
induces non-equilibrium weak reactions, which release energy that
is invested partly in neutrino emission and partly in internal
heating. Eventually, the star arrives at a stationary state in
which the temperature remains nearly constant, as the forcing
through the change of $G$ is balanced by the ongoing reactions.
Using the surface temperature of the nearest millisecond pulsar
(PSR J0437$-$4715) inferred from ultraviolet observations and
results from theoretical modelling of the thermal evolution, we
estimate two upper limits for this variation: (1) $|\dot G/G| < 2
\times 10^{-10}~\mathrm{yr}^{-1},$ if the fast, ``direct Urca''
reactions are allowed, and (2) $|\dot G/G|<4\times
10^{-12}~\mathrm{yr}^{-1},$ considering only the slower,
``modified Urca'' reactions. The latter is among the most
restrictive upper limits obtained by other methods.

 }

\maketitle{}

\section{Introduction}

A number of theorists have proposed that the so-called
``fundamental constants" of Nature may vary with cosmological
time, ever since Dirac suggested that the gravitational force may
be weakening \citep{dirac37}. Several experiments aimed at
constraining the time variation of $G$ have been conducted, and
their results can be grouped by the time scales they probe (see
\citealt{reisenegger07} for a list with references): (1) \emph{Human
lifetime} ($\sim 10$~yr) experiments rely on real-time monitoring
of distances within the Solar System, white-dwarf oscillation
periods, and pulse arrival times of isolated and binary pulsars,
(2) \emph{long time} ($\sim 10^{9-10}$~yr) measurements use
stellar astrophysics and paleontology, and (3) \emph{cosmological}
($\gtrsim 10^{10}$~yr) experiments use the Big Bang
nucleosynthesis and Cosmic Microwave Background anisotropies to
compare the value of $G$ in the early Universe to the present
value. Generically, they set constraints on $|\dot{G}/G|$ down to
$\sim 10^{-12}$~yr$^{-1}$, although the comparison between
different timescales  depends on the assumed form of the function
$G(t)$.

Here, we review our previously introduced method for setting
constraints on $\dot{G}$ \citep{jofre06}, to our knowledge the
only one so far to probe timescales $\sim 10^{7-9}~\mathrm{yr}$,
intermediate between the \emph{human} and \emph{long} timescales
mentioned above. It relies on the change in the internal structure
of a neutron star induced by a change in $G$, which, together with
the slow response timescale of weak interactions, result in
internal heating and an increase in the observed surface
temperature.

Neutron stars have mean densities exceeding nuclear saturation
density, $\rho_{\rm nuc}\sim 3\times 10^{14}$~g~cm$^{-3}$. In
their outer layers, they are composed of heavy atomic nuclei and
free electrons, giving way to free neutrons, free protons, muons,
and potentially more exotic particles as the density increases
inward. A short time after their formation, their internal
temperatures drop orders of magnitude below the Fermi energies of
free particles ($\sim 10-100$~MeV), and so their structure is well
approximated by zero-temperature models.

Nonetheless, weak interactions still play an important role, as
characteristic temperatures $10^{6-8}$~K yield non-negligible
neutrino emission. Given that the equation of state of matter
above nuclear density still remains poorly known, researchers
construct models to calculate the evolution of the thermal content
of neutron stars and compare their predictions with X-ray
observations, in order to set constraints on the models for dense
matter (e.g., \citealt{yakovlev04}). In these thermal evolution
models, the structure of the star is calculated assuming an
equation of state, and the evolution of the internal temperature
is obtained by considering losses due to different neutrino
emission processes from the stellar interior, as well as thermal
electromagnetic radiation from the surface.

We have previously explored \emph{rotochemical heating}, the
effect of the progressive loss of rotational support of
millisecond pulsars has on their internal structure, which leads
to weak interaction processes (beta decays) and net heating
\citep{reisenegger95,reisenegger97,fernandez05}. A time variation
of the gravitational constant has an analogous effect: it
compresses the star and therefore also causes heating
(\emph{gravitochemical heating}, \citealt{jofre06}). In what follows,
we discuss how this process constrains $\dot{G}$.

%------------------------------------------------------------------
\section{Method}

Neutron star matter is composed of degenerate fermions of various
kinds: neutrons ($n$), protons ($p$), leptons ($l$), and possibly
other, more exotic particles. Neutrons are stabilized by the
presence of other, stable fermions that block (through the Pauli
exclusion principle) most of the final states of the beta decay
reaction $n\to p+l+\bar\nu$, making it much slower than in vacuum.
The large chemical potentials $\mu_i$ ($\approx$ Fermi energies)
for all particle species $i$ also make inverse beta decays,
$p+l\to n+\nu$, possible. The neutrinos ($\nu$) and antineutrinos
($\bar\nu$) leave the star without further interactions,
contributing to its cooling. The two reactions mentioned tend to
drive the matter into a chemical equilibrium state, defined by
$\eta_{npl}\equiv\mu_n-\mu_p-\mu_l=0$. Depending on the relative
abundance of protons, the reactions mentioned above (called
\emph{direct Urca processes}) may not be able to conserve
momentum, in which case an additional nucleon must participate in
the reactions (yielding the \emph{modified Urca processes}, e.~g.,
\citealt{yakovlev01}), with strongly reduced rates.

If $G$ changes, so does the hydrostatic equilibrium structure of
cold neutron stars, and all matter elements in their interior are
compressed or expanded, and in this way driven away from chemical
equilibrium ($\eta_{npl}\neq 0$). Free energy is stored, which is
released by an excess rate of one reaction over the other.
This energy is partly lost to neutrinos and antineutrinos
(undetectable at present), the remainder heats the stellar
interior. The heat is eventually lost as thermal (typically
ultraviolet) photons emitted from the stellar surface.

The thermal evolution of the star is calculated by solving a system
of ordinary differential equations of the form
\begin{eqnarray}
\label{eq:Tdot}
\dot{T} & = & \frac{1}{C(T)}\left[L_H(T,\eta_{npl}) - L_\nu(T,\eta_{npl})\right.\nonumber\\
        &   & \left.\phantom{C(T)\,L_H(T,\eta_{npl})- L_\nu} - L_\gamma(T)\right]\\
\label{eq:etadot} \dot{\eta}_{npl} & = &A(T,\eta_{npl}) + B\,
\dot{G},
\end{eqnarray}
where $T$ is the internal temperature, $C$ the heat capacity of
the star, $L_H$, $L_\nu$, and $L_\gamma$ the heating, neutrino
cooling, and photon cooling luminosities, respectively, $A$ represents the
change in the particle abundances due to reactions, and $B$ a
constant coefficient that depends only on the structure of the
star. See \cite{jofre06} and \cite{fernandez05} for more details.

%--------------------------------------------------------------------
\section{Results}

The typical result of integrating equations~(\ref{eq:Tdot}) and
(\ref{eq:etadot}) is as follows (details for the analogous,
rotochemical heating case are given in \citealt{fernandez05} and
\citealt{reisenegger07}). The star first cools down (within $\lesssim
10^7~\mathrm{yr}$) from its high birth temperature, while the
chemical potential imbalances $\eta_{npl}$ slowly increase due to
the gravitational forcing term in equation~(\ref{eq:etadot}).
After $\sim 10^{7-9}~\mathrm{yr}$, depending on $|\dot G|$, the
imbalances are so high that they increase the reaction rates to
the point where they keep up with the forcing, stabilizing the
chemical imbalance. At this point, the energy released by
reactions and that emitted from the surface also compensate in
equation~(\ref{eq:Tdot}), keeping the temperature at a constant
value $\sim 10^5~\mathrm{K}$, whose precise value can be predicted
if $|\dot G|$ and the neutron star model (mass and equation of
state) are given.

\begin{figure}
\centering
\includegraphics[scale=0.4]{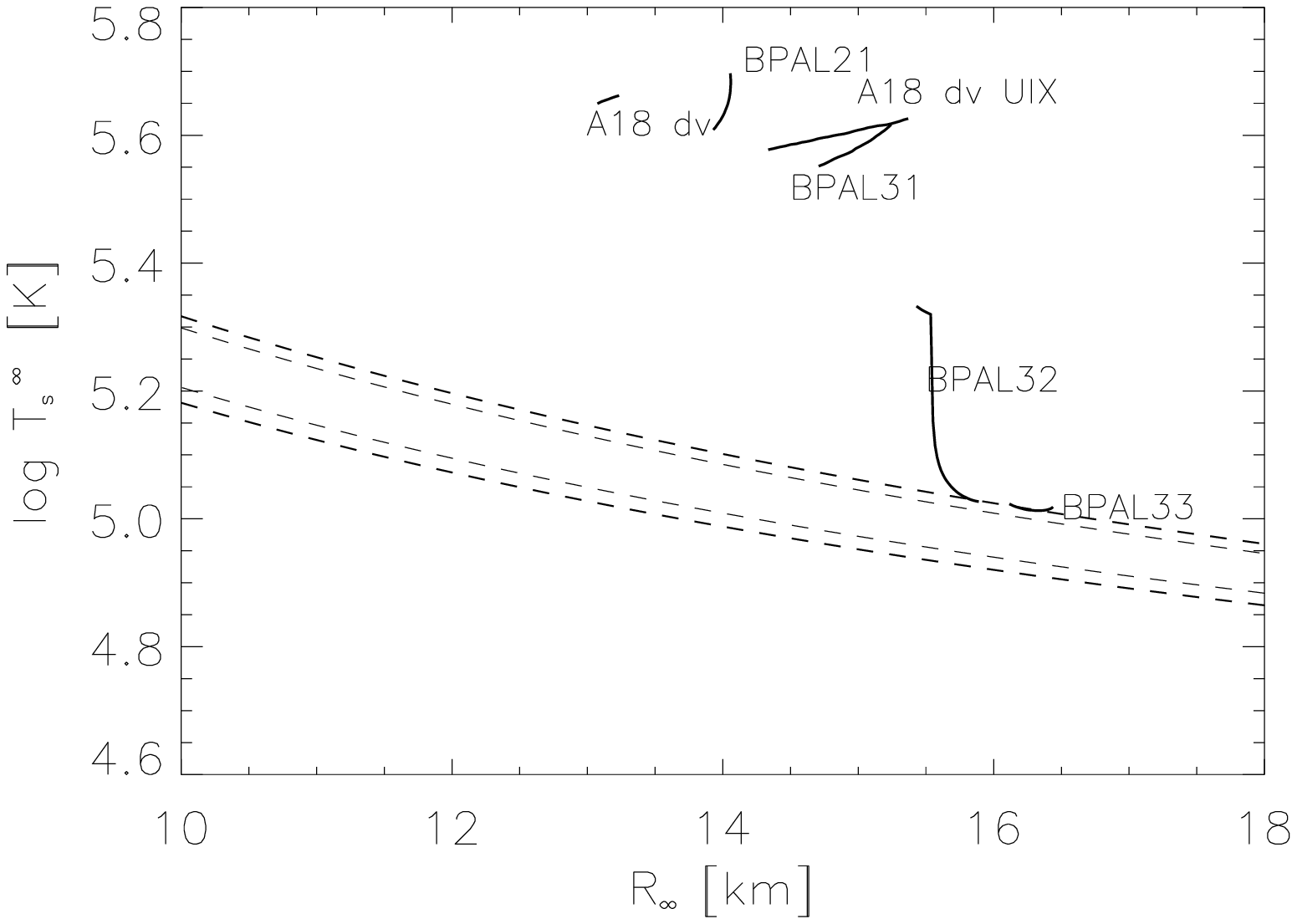}
\caption{Comparison of the gravitochemical heating predictions to
observations of PSR~J0437-4715. The solid lines are the predicted
stationary surface temperatures as functions of stellar radius,
for different equations of state (A18 from \citealt{apr98} and BPAL
from \citealt{pal88}), constrained to the observationally allowed
mass range for this pulsar~\cite{vbb01}. Dashed lines correspond
to the 68\% and 90\% confidence contours of the blackbody fit of
\cite{kargaltsev04} for the ultraviolet emission
from this object. The value of $|\dot{G}/G| = 2 \times
10^{-10}$~yr$^{-1}$ is chosen so that all stationary temperature
curves lie above the observational constraints. (BPAL32 and BPAL33
allow direct Urca reactions in the observed mass range of J0437.)}
\label{J0437_1}
\end{figure}

\begin{figure}
\centering
\includegraphics[scale=0.4]{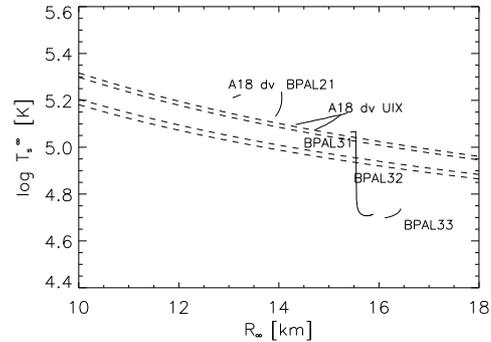}
\caption{Same as Figure~\ref{J0437_1}, but now the value of
$|\dot{G}/G| = 4 \times 10^{-12}$ yr$^{-1}$ is chosen so that only
the stationary temperature curves with modified Urca reactions are
above the observational constraints.}
\label{J0437_2}
\end{figure}

In order to constrain $\dot G$, we therefore need sensitive
ultraviolet observations of known neutron stars older than
$10^7~\mathrm{yr}$. Such an observation only exists for the
nearest millisecond pulsar, PSR J0437$-$4715, for which an HST
observation detected what appears to be the ultraviolet
Rayleigh-Jeans tail of a blackbody at $\sim 10^5~\mathrm{K}$
\citep{kargaltsev04}. Figures~\ref{J0437_1} and \ref{J0437_2}
compare theoretical predictions for various equations of state and
a range of neutron star masses to the error contours obtained from
this observation. The upper limit on $|\dot{G}/G| = 2 \times
10^{-10}$~yr$^{-1}$ (Fig.~\ref{J0437_1}) is obtained by requiring
that all stationary temperature curves lie above the observational
constraints. Restricting the models to only modified Urca
reactions yields a more restrictive constraint, $|\dot{G}/G| = 4
\times 10^{-12}$ yr$^{-1}$ (Fig.~\ref{J0437_2}).

\section{Discussion}

\emph{Gravitochemical heating} sets constraints on the time
variation of the gravitational constant, using the fact that a
non-zero change would generate internal heating in neutron stars,
which for nearby cases can be detected with existing telescopes.
These constraints are the only ones on timescales $10^{7-9}$~yr.
If it could be assured that the observed neutron stars cannot have
direct Urca reactions, these constraints would be of the same
order as the best ones available on other time scales. This is not
the case at the moment, but could become so as neutron star
interiors become better understood from other studies. In
particular, observed temperatures of other old neutron stars (such as
millisecond pulsars) will provide useful information.

\begin{acknowledgements}
This work was supported by FONDECYT Regular Grant 1060644, the
FONDAP Center for Astrophysics, and the Basal Funding Project
PFB-06/2007 (\emph{Center for Astrophysics and Related
Technologies}).
\end{acknowledgements}

%---------------------------------------------------------------
\bibliographystyle{aa}

\end{document}